\documentclass[
reprint,
superscriptaddress,
 amsmath,
 amssymb,
 aps,
 pre,
 floatfix,
]{revtex4-2}

\usepackage{graphicx}
\usepackage{bm}
\usepackage{xcolor}
\usepackage{hyperref}
\usepackage{tasks}

\settasks{style=itemize, item-format={\scshape\small}, column-sep={-1mm}, after-item-skip=-1mm, after-skip={2mm}}

\usepackage[bf]{subfigure}

\begin{document}

\preprint{APS/123-QED}

\title{On the Stability of Citation Networks}

\author{Alexandre Benatti}
\author{Henrique Ferraz de Arruda}
\affiliation{
 S\~ao Carlos Institute of Physics,
University of S\~ao Paulo, PO Box 369,
13560-970, S\~ao Carlos, SP, Brazil 
}

\author{Filipi Nascimento Silva}
\affiliation{Indiana University Network Science Institute, Bloomington, IN, USA
}

\author{C\'esar H. Comin}
\affiliation{Department of Computer Science, Federal University of S\~ao Carlos, S\~ao Carlos, Brazil
}

\author{Luciano da Fontoura Costa}
\affiliation{
 S\~ao Carlos Institute of Physics,
University of S\~ao Paulo, PO Box 369,
13560-970, S\~ao Carlos, SP, Brazil 
}


\date{\today}

\begin{abstract}
Citation networks can reveal many important information regarding the development of science and the relationship between different areas of knowledge. Thus, many studies have analyzed the topological properties of such networks. Frequently, citation networks are created using articles acquired from a set of relevant keywords or queries. Here, we study the robustness of citation networks with regards to the keywords that were used for collecting the respective articles. A perturbation approach is proposed, in which the influence of missing keywords on the topology and community structure of citation networks is quantified. In addition, the relationship between keywords and the community structure of citation networks is studied using networks generated from a simple model. We find that, owing to its highly modular structure, the community structure of citation networks tends to be preserved even when many relevant keywords are left out. Furthermore, the proposed model can reflect the impact of missing keywords on different situations.
\end{abstract}

\maketitle


\section{\label{sec:introduction}Introduction}

The study of how science evolves is essential not only as a means to understand how knowledge is organized, but also as a subsidy for devising ways to help scientists while accessing research topics and even to optimize the scientific advancement~\cite{de2017knowledge}. The field that deals with this type of study is called Science of Science (SciSci)~\cite{fortunato2018science}. The recent increase in the availability of organized datasets and computational power gave rise to many systematic and large-scale studies in SciSci. Some examples include understanding how scientific fields are connected~\cite{de2018integrated}, the evolution of scientific interest over time~\cite{aleta2019explore}, citation patterns arisen from social media~\cite{thelwall2013altmetrics, erdt2016altmetrics}, predicting the future collaborations of authors ~\cite{tuninetti2020prediction}, among other possibilities.

A promising approach to SciSci consists in obtaining networks representing the organization of scientific and technological knowledge that can be drawn from the citations among documents, such as in an encyclopedia, body of publications, patents, etc.  These networks are particularly interesting because they allow, among other possibilities, the identification of scientific areas and sub-areas corresponding to respectively identified clusters and communities.  In addition, these networks can also provide insights into the hierarchical organization of scientific areas and even research groups.  

Many studies have approached the analysis of citation networks~\cite{fortunato2018science}, with some of them focusing on the relationship between areas of science~\cite{meyer2010can,tussen2000technological,shibata2010extracting,gazis1979influence}.
These analyses are now commonly undertaken on large datasets~\cite{rosvall2010mapping}, which only recently became available. This includes the MAG (Microsoft Academic Graph)~\cite{sinha2015overview}, APS (American Physical Society)~\cite{APS2021}, PLOS journals~\cite{PlosOne}, Wikipedia~\cite{wiki2021}, and ArXiv~\cite{arxiv2021}. Also, much of this type of research deals with the detection and investigation of groups of papers that are strongly related, forming network communities, which can also be analyzed in a multi-scale fashion~\cite{rosvall2011multilevel}, as well as their evolution over time~\cite{rosvall2010mapping,lancichinetti2012consensus}. For instance, in~\cite{lancichinetti2012consensus} a consensus community approach is proposed, which has been employed in the temporal analysis of citation networks. 

\begin{figure*}[!ht]
  \centering
    \subfigure{\includegraphics[width=.9\textwidth]{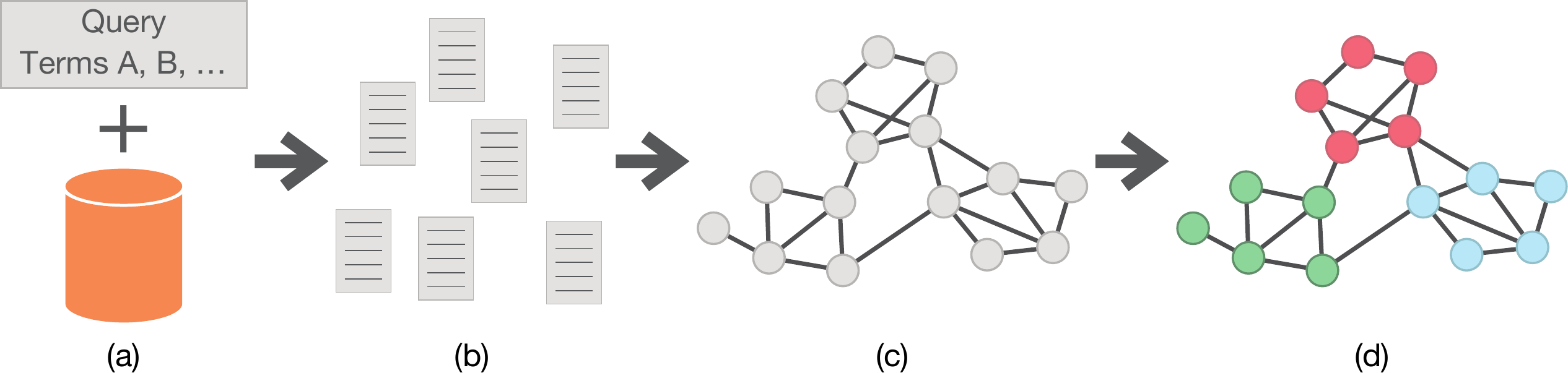}}
  \caption{The basic framework underlying the derivation of a citation network. (a) It starts with a specific query on a database; (b) the recovered documents are selected; (c) from the documents and their citations, a network is created; (d) the respective communities are identified, and labels are attributed to it.}
  \label{fig:scheme_summary}
\end{figure*}

In~\cite{silva2016using} the authors proposed a methodology that was able to identify keywords that label the detected communities of citation networks. More specifically, the authors proposed a pipeline that starts from the data acquisition to community detection~\cite{fortunato2010community} and its respective labeling procedure, which is illustrated in  Figure~\ref{fig:scheme_summary}. In such a type of work, queries are used to find a core set of papers related to a certain topic, which in turn are used to construct a citation network. For example, in ~\cite{ceribeli2021coupled} the authors analyzed two subareas of chemistry and found that some communities are potentially more related to the chemical methods while others to applications. In another work, the authors investigated a way to enrich citation networks that ended up being too small by expanding the core set of papers based on related cited materials~\cite{benatti2021enriching}.

Though founded on a sound logical framework, some important issues remain regarding the extraction of citation networks for specific topics.  Of particular importance is how much the obtained networks can vary as a consequence of distinct, though related, choices of adopted queries. In other words, if markedly distinct networks are obtained by relatively minor query changes, they will have limited value regarding the respective analysis from the perspective of science of science.  Therefore, a more systematic investigation of the stability of citation networks given variations of related query terms constitutes an important aspect deserving further attention. The present work reports an approach focusing on this important question. In particular for networks obtained from core sets of papers that were matched against a set of keywords in a large dataset.

Two complementary methods were regarded to tackle this problem. First, we performed an analysis of the overall citation network in terms of some particularly important topological measurements without focusing on the respective community structure. Given its special importance for scientific interpretations, the modular organization of the obtained structures was specifically analyzed as a complementary study. More specifically, we considered measurements aimed at quantifying the quality of the obtained clusters.  In addition, we took into account two distinct strategies for node removal, aimed at keeping the network as small and as large as possible, respectively. A toy model, representing some of the typical structure of citation networks, was also employed to complement our stability investigation.

Many interesting results were obtained. First, in the analysis of the network topology, we compared the modified networks, which are versions created by removing a single keyword. For some specific cases, the topology was found to be strongly affected. For many sets of modified networks, the organization was not found to change considerably. We also analyzed the scenario in which keywords were successively removed. As a conclusion, we observed that independently of the deletion strategy, it is necessary to remove many keywords to significantly affect the community structure of the networks. To complement this analysis, we considered a toy model. By varying the toy model parameters, we could then test parameters that are present in the employed network and other possibilities of organization. More specifically, with this model, we can control both the network topology and how the keywords are distributed among communities.  

The sections of this paper are organized as follows. Section~\ref{sec:method} presents the characteristics of the employed dataset, how the networks are created, the robustness analysis, and the adopted toy model. In Section~\ref{sec:results}, we present the obtained results and the respective discussions. The conclusions are presented in Section~\ref{sec:conc}.

\section{Data collection and analysis}\label{sec:method}

In this section, we describe the employed dataset, as well as the proposed methodologies for the network analysis, from the topological measurements to the different strategies of robustness based on the keywords.

\subsection{Dataset}

The articles used in our analysis were obtained from the Microsoft Academic Graph (MAG)~\cite{sinha2015overview}, which is a dataset that contains scientific publication records, citations between those publications, and other information. More information is described on MAG's website\footnote{\url{https://www.microsoft.com/en-us/research/project/microsoft-academic-graph/}}. The employed raw data MAG was provided by the Collaborative Archive \& Data Research Environment (CADRE) project at Indiana University~\citep{mabry2020cadre}.

In this work, we employed the data set with publications up to 25th June, 2020. Then, we searched for some keywords that represent a given area of interest. These keywords were selected based on the methods listed in the Wikipedia page of Pattern recognition\footnote{\url{https:\/\/en.wikipedia.org\/wiki\/Pattern_recognition}, accessed on 30 October 2020.}. Here, we considered the following keywords:

\begin{tasks}(2)
    \task \emph{Linear discriminant analysis}
    \task \emph{Quadratic discriminant analysis}
    \task \emph{Maximum entropy classifier}
    \task \emph{Decision tree}
    \task \emph{Decision list}
    \task \emph{Kernel estimation}
    \task \emph{K-nearest-neighbor}
    \task \emph{K-nearest neighbor}
    \task \emph{K nearest neighbor}
    \task \emph{Naive Bayes classifier}
    \task \emph{Neural network}
    \task \emph{Perceptron}
    \task \emph{Support vector machine}
    \task \emph{Gene expression programming}
    \task \emph{Categorical mixture model}
    \task \emph{Hierarchical clustering}
    \task \emph{K-means clustering}
    \task \emph{Correlation clustering}
    \task \emph{Kernel PCA}
    \task \emph{Boosting}
    \task \emph{Bootstrap aggregating}
    \task \emph{Ensemble averaging}
    \task \emph{Mixture of expert}
    \task \emph{Bayesian network}
    \task \emph{Markov random field}
    \task \emph{Kalman filter}
    \task \emph{Particle filter}
    \task \emph{Gaussian process regression}
    \task \emph{Kriging}
    \task \emph{Linear regression}
    \task \emph{Independent component analysis}
    \task \emph{Principal component analysis}
    \task \emph{Conditional random field}
    \task \emph{Hidden Markov model}
    \task \emph{Maximum entropy}
    \task \emph{Markov model}
    \task \emph{Hidden Markov model}
    \task \emph{Dynamic time warping}
\end{tasks}

In order to generate citation networks, we considered all titles and abstracts that have at least one of the predefined keywords. We considered all citations between these selected documents. We also removed non-connected documents and considered only the largest connected component of the network. 

\subsection{Topology analysis}
\label{sec:met_topology}

One approach for comparing networks is by using a set of respective measurements. Many distinct types of measurements have been proposed in the literature~\cite{costa2007characterization,boccaletti2006complex}. Because of the large size of our analyzed network, we employed only a set of local measures that do not demand much computational power. The employed measurements are listed as follows:
\begin{itemize}
  \item \emph{Degree}: For the undirected case, the degree is defined as the number of edges connected to a given node ($k_{all}$). For directed networks, there are two definitions the consider the counts of incoming ($k_{in}$) and outgoing edges ($k_{out}$). We considered $k_{all}$, $k_{in}$, and $k_{out}$ as three separated measurements.
  \item \emph{Clustering coefficient}~\citep{watts1998collective}: since there are different definitions of clustering coefficient for directed networks, we employed the measurement for the undirected case. The undirected version of this measurement is defined with bases on the number of connected triples (triangles), $\Delta(i)$, and the number of edges, $\tau(i)$, as $c_i = \frac{\Delta(i)}{\tau(i)}$;
 \item \emph{Network size}: Number of network nodes;
 \item \emph{Average degree of the neighbors}: For each network node, the neighbors are selected and the measurement is computed as the average degree of these selected nodes.
\end{itemize}
Because the employed measurements are defined to be calculated for network nodes, we considered the average of these measurements to compare the networks. 

\subsection{Robustness analysis}
\label{sec:robust}

Here, we propose a technique for estimating the robustness of a network. This technique consists of comparing the community structure obtained from the networks created by considering all keywords and \emph{modified versions} of this network, which consists of networks created from a respective subset. We opted for this approach because the choice of keywords depends on the researcher that builds the network. More details regarding the strategies employed to create modified versions of the network are described in Section~\ref{sec:res_comparison}. 

First, it is necessary to choose a method for community detection. It is important to point out that a similar analysis could be developed by considering different community detection methods. There are many possibilities of algorithms devoted to identifying communities, also called clusters, in networks~\cite{fortunato2010community}. In this study, we considered \emph{Infomap}~\citep{rosvall2009map} that has been employed in related applications of the science of science studies~\citep{rosvall2008maps,rosvall2010mapping}. This approach was also used in~\citep{benatti2021enriching,ceribeli2021coupled}, as the step of paper clustering before computing the respective labels.   Also, the citation networks can have clusters with few nodes. Thus, it is important to avoid using methods of community detection based on modularity~\cite{newman2006modularity} because of the resolution limit~\cite{fortunato2007resolution}, which can merge small communities.

\begin{figure*}[!ht]
  \centering
    \subfigure{\includegraphics[width=0.99\textwidth]{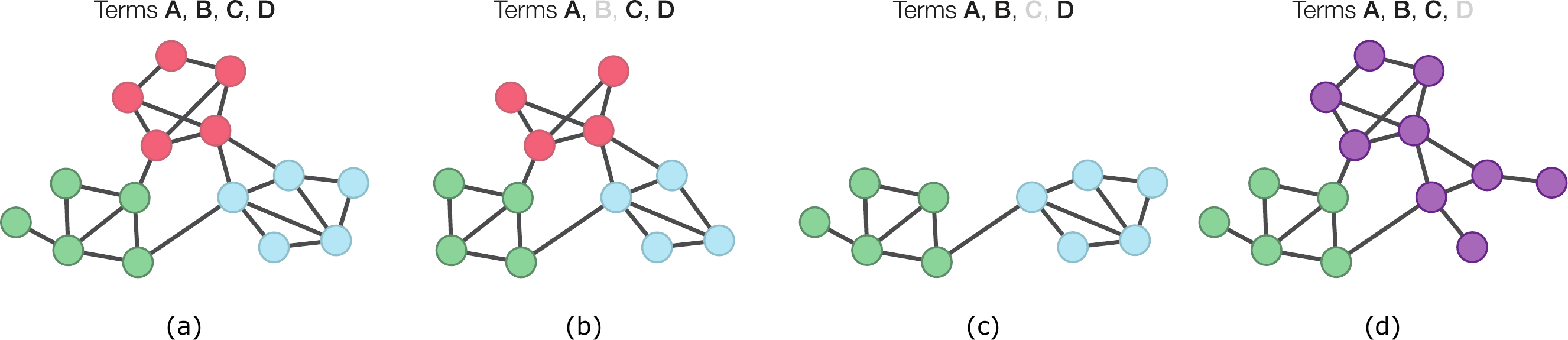}}
  \caption{Example of possibilities of modified networks, in which the colors represent the identified node communities. (a) represents the original network, and (b), (c), and (d) are three possible modified versions. In (b), though a node was deleted, the original communities were kept. In (c), an entire community was deleted but the other communities were kept. In (d), the removal of keyword D implied the merging of two communities.}
  \label{fig:robust}
\end{figure*}

In more detail, Infomap is related to two concepts: random walks and Huffman code~\cite{huffman1952method}. The random walk is performed on the network. This algorithm considers that the description can be minimized for networks with communities. The description lengths of the codes (binary numbers) are optimized in order to minimize the sizes of the trajectory lengths. This hypothesis reflects the fact that the random walker tends to visit nodes from the same region for many consecutive iterations. The random walk dynamics, as well as the optimization, are executed for $I$ iterations, and the detected communities are considered for the case in which the best optimization is found. More details regarding this method are described in~\cite{bohlin2014community}. We considered the optimization of ``two-level'', $I=1000$, and the undirected version. The detected communities are stored in a vector, where each position represents a network node. Figure~\ref{fig:robust} illustrates examples of modified networks as well as the detected communities.

The next step is to compare the communities identified in the original network and the respectively obtained modified versions.  Observe that the latter can have different numbers of nodes. In this case, we compared only the community vector positions that take part in both networks. Note that the indices that represent communities can vary. For example, if the list of communities of the networks $G_1$ and $G_2$ are $c_1=\{a,a,b,b,b\}$ and $c_2=\{c,c,d,d,d\}$, respectively, the detected communities are exactly the same but with distinct indices. The community $a$ for $c_1$ is equivalent to $c$ in $c_2$ and  $b$ for $c_1$ is equivalent to $d$ in $c_2$. Consequently, in this case, a given measurement of comparison between this pair of lists would return the maximum value, representing a perfect match. To overcome this problem, we tested measurements proposed to account for the quality of clusters by considering the match between a pair of lists, as follows:

\begin{itemize}
    \item \emph{Normalized Mutual Information} (NMI)~\cite{kuncheva2004using}: This measurement is calculated in terms of the mutual information (MI), a concept proposed in the information theory area. More specifically, by considering the Shannon entropy, MI measures how much information of one variable is expressed in another variable. Normalized Mutual Information (NMI) is a MI normalization. There are some distinct possibilities of normalization. Here we considered entropy averages as the normalization parameter. This measurement is commonly used to deal with problems related to community detection~\cite{lancichinetti2008benchmark};
    
    \item \emph{Adjusted Mutual Information} (AMI)~\cite{vinh2010information}: As well as in NMI, AMI is based on the concept of mutual information to compare clusters, although this method proposes to decrease the effect of concordance due to chance. AMI is performed by calculating the mutual information (MI) and then subtracting the value of the expected mutual information ($E\{\text{MI}\}$), which is determined by considering two random clusters;  
    
    \item \emph{Adjusted Rand Index} (ARI)~\cite{hubert1985comparing}: ARI is the corrected-for-chance version of the rand index (RI), which is defined as the ratio between the number of agreements (true positives plus true negatives) and the total number of pairs. The RI correction (ARI) is based on the expected similarity of the comparison between all pairs generated by a random model;
    
    \item \emph{V-measure} (VME)~\cite{rosenberg2007v}: The V-measure is also an entropy-based measurement that expresses how successfully the criteria of homogeneity and completeness have been satisfied, estimating these measures weighted harmonic mean. The weighting is performed by a parameter $\beta$. If $\beta < 1$ homogeneity is more strongly weighted and if $\beta > 1$ completeness is more strongly weighted, and for the case in which that $\beta=1$, this measurement recovers the measurement of NMI.
\end{itemize}

The values obtained with these measures are normalized between 0 and 1. More specifically, 0 and 1 represents that the variables are completely independent and dependent, respectively. The functions employed for community comparison were implemented by \emph{sklearn.metrics} available in the \emph{scikit-learn} library~\cite{scikit-learn}.

\subsection{Toy model}
\label{sec:toy}

To test different possibilities of analysis with a given set of network features, we proposed the analysis on a network model of communities. More specifically, we considered the LFR-benchmark~\cite{lancichinetti2008benchmark}, which is a model that generates networks with communities and power-law degree distribution. In this network model, it is possible to control the percentage of edges connecting different communities through the mixing parameter, $\mu$, the exponent of the power-law distribution, $\gamma$, the min and max community sizes ($S_{min}$ and $S_{max}$),  and the maximum node degrees ($k_max$). There is also the possibility of setting the number of overlapping nodes and the number of these nodes' memberships. We set both these values to zero. The remaining adopted parameters are described in Section~\ref{sec:res_toy}.

In order to compare different executions of our analysis, for each of the employed parameters set, we generate networks with the same number of communities $C$. Then, we obtain many distinct networks and selected only the samples with the fixed number of communities $C$.

Two strategies of assigning artificial keywords were proposed: 
\begin{itemize}
\item \emph{Keywords dependent of communities}: we considered that each community represents a given keyword. More specifically, the keywords vector is a copy of the community membership vector. Consequently, all the nodes representing a community are removed when a given keyword is left out;
\item \emph{Keywords independent of communities}: we also investigate if the keywords could not be associated with the communities. To test this hypothesis, we considered the keywords as a shuffled version of the keywords vector proposed in the \emph{Keywords dependent of communities} strategy.
\end{itemize}

\section{Results and discussions}\label{sec:results}

We start our analysis by characterizing and comparing the topology of the different modified networks based on the area of \emph{pattern recognition}. In the following, we tested the different deletion strategies of nodes and quantified the results through measurements of the clusters' quality. To better understand the dynamics involved in this process, we also considered a toy model, which is analyzed using the same methodology employed in the other subsections.

\subsection{Network characterization}
\label{sec:characterization}

\begin{figure}[!ht]
  \centering
    \subfigure{\includegraphics[width=0.45\textwidth]{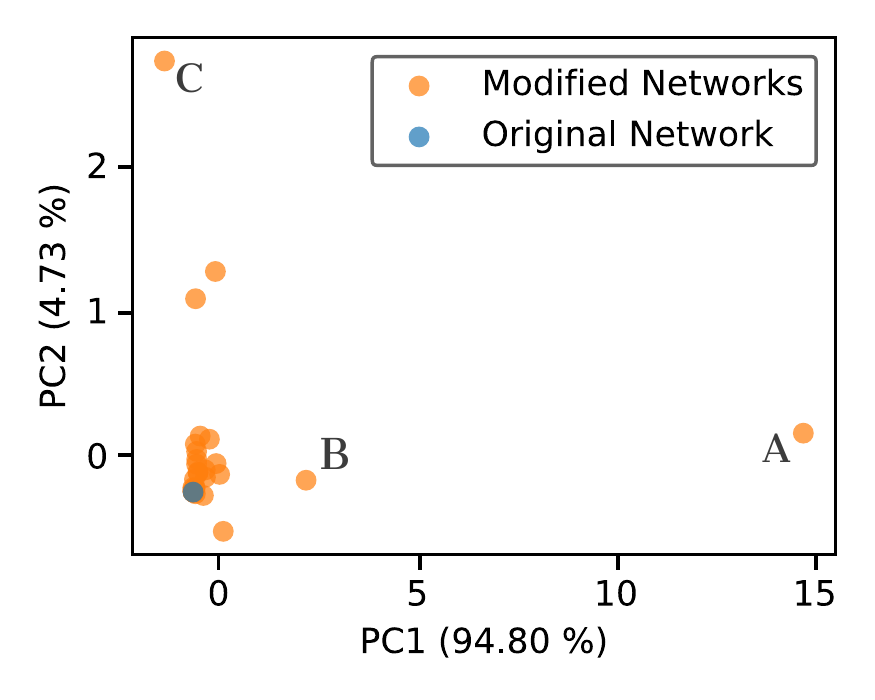}}
  \caption{Principal component analysis projection considering the employed network measurements. The original and all modified network versions were considered. The modified versions represent the deletion of: A - \emph{Neural network},  B - \emph{Support vector machine}, and  C - \emph{Linear regression}.}
  \label{fig:topology}
\end{figure}

First of all, we create the network by considering all the selected keywords. The resulting network has $897,991$ nodes, and an average degree of $11.93$. In order to characterize and compare the different versions of the modified network, we employed the network measurements described in Section~\ref{sec:met_topology}. Due to this network's large size, we considered only measurements that do not demand a considerable processing time. We projected the data by using a Principal Component Analysis (PCA)~\cite{gewers2018principal,jolliffe2016principal} (see Figure~\ref{fig:topology}). The first principal component (PC1) represents 94.8\% of the data variation. So, our analysis is based manly on this axis. 

For most modified networks, the topology tends to result similar to that of the original network. This result is obtained because, for the majority of the cases, the network size is not considerably changed. However, for the case in which we removed documents obtained by deleting the keyword \emph{Neural network}, the resultant modified network is the most different. This is the smaller network, with the lowest average degree. See point A on the right-hand side of the PC1 axis of Figure~\ref{fig:topology}. The second most distant to the original network, in PC1, is the modified network considering the deletion of nodes with the keyword \emph{Support vector machine} (see point B in Figure~\ref{fig:topology}). 
In this case, the average degree is the second-lowest. In contrast, by considering the deletion of nodes obtained from the word \emph{Linear regression}, we obtained the second smallest network. It is indicated as C in Figure~\ref{fig:topology}. However, it was found to be on the opposite side of the case of \emph{Neural network}, which followed because the network created without \emph{Linear regression} has the highest average degree. More details regarding the different network sizes are described in Section~\ref{sec:res_comparison}. 

\subsection{Comparison between original and modified networks}
\label{sec:res_comparison}

As well as in Section~\ref{sec:characterization}, we start our analysis by considering the modified networks as being created from all original keywords, except for a given keyword. For both the original and modified networks, the communities were calculated using the following parameters of Infomap: (i) \emph{two-level} and (ii) number of iterations 1000. We empirically observed that a high number of iterations is crucial to return consistent results. This parameter is essential due to the large size of the original network. In the following analyses, we employed the same set of parameters presented here. 

\begin{figure*}[!ht]
  \centering
    \subfigure{\includegraphics[width=1.\textwidth]{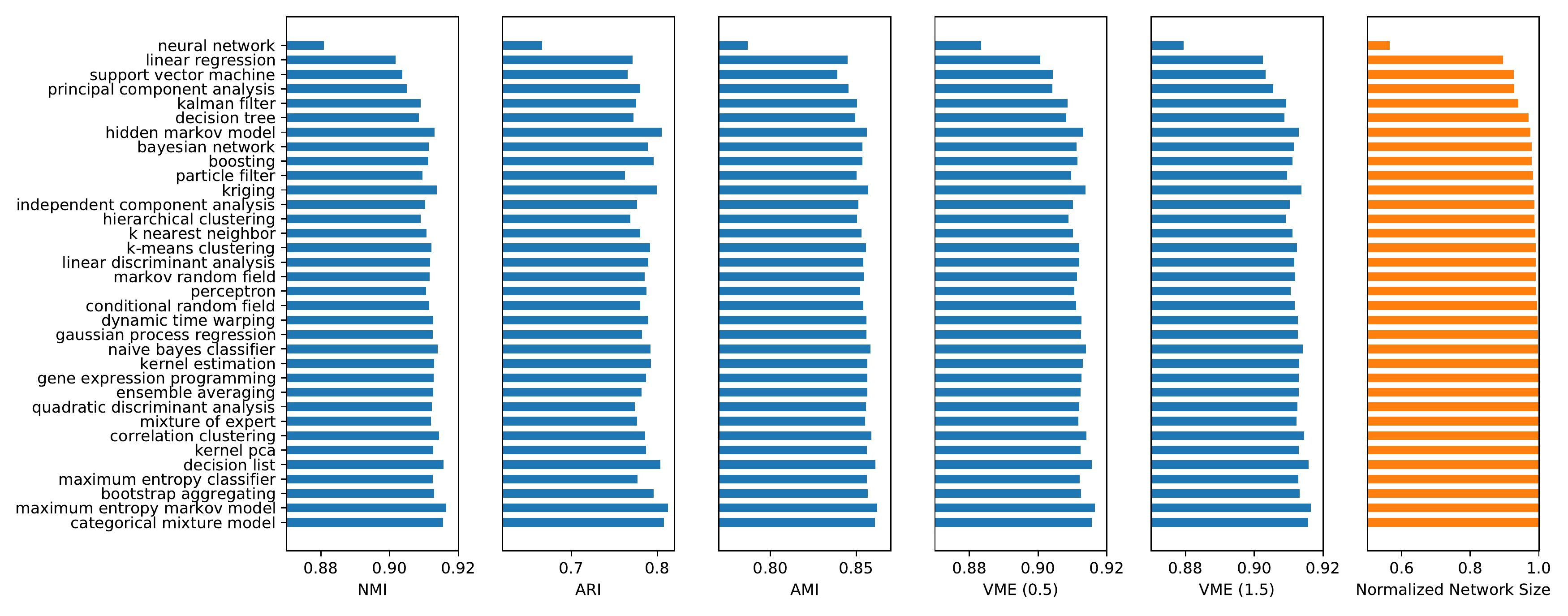}}
  \caption{Comparison among different quality measurements of the clusters. All the clusters are obtained by considering the modified networks with a single keyword deletion. Shown in the plots are the keywords removed. The subplot in orange shows the normalized sizes of the modified networks.}
  \label{fig:NMI_ARI_AMI_VME_sizes}
\end{figure*}

As the first analysis, regarding the differences between modified networks and the detected communities, we considered all the measurements of quality of clusters, described in Section~\ref{sec:robust} (see Figure~\ref{fig:NMI_ARI_AMI_VME_sizes} illustrates). For all measurements, the results were found to be similar. Interestingly, the modified networks without the words \emph{Neural network}, \emph{Linear regression}, and \emph{Support vector machine} tend to have lower values of agreement between their communities and the original network. As well as in the analysis of the network topology, this result follows from the fact that the networks have fewer nodes than the other versions of modified networks. Moreover, this result is more evident when the network created without the word \emph{Neural network} is considered. 

\begin{figure}[!ht]
  \centering
    \subfigure{\includegraphics[width=0.5\textwidth]{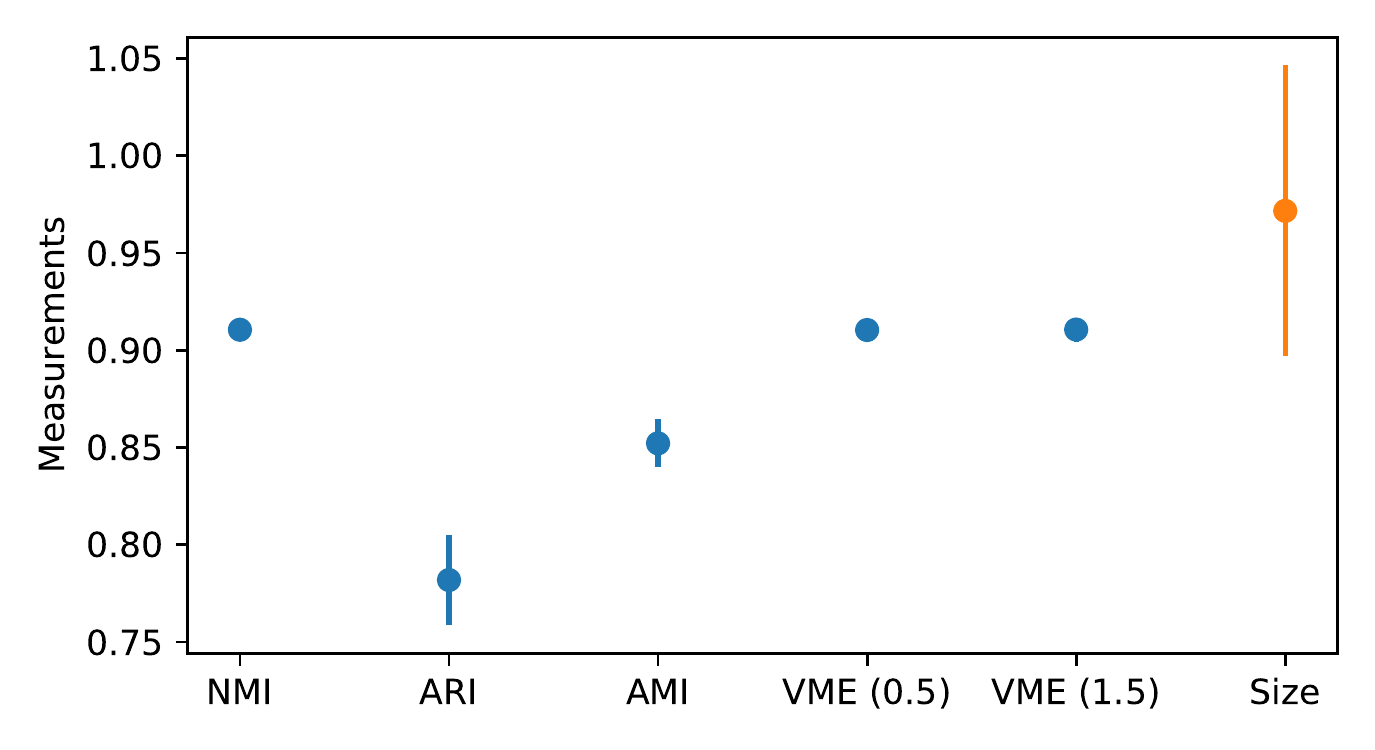}}
  \caption{Average and the standard deviation of measures presented in Figure~\ref{fig:NMI_ARI_AMI_VME_sizes}.}
  \label{fig:measures}
\end{figure}

The modified network sizes were normalized by the number of nodes of the original network, as shown in Figure~\ref{fig:NMI_ARI_AMI_VME_sizes}. In order to better understand if the variation of the community structures and network sizes are comparable, we obtain their standard deviations, as shown in Figure~\ref{fig:measures}. Interestingly, for all of the quality measurements of the clusters, the respective standard deviations are substantially lower than for the normalized network size. Thus, this result indicates that the employed measurements can be more dependent on the community organizations than the network size.  This suggests that the communities will tend to be preserved with the removal of keywords.

For all the employed measurements, the results were found to be similar. So, in the remainder of this study, we considered only the NMI measurement. Furthermore, for all modified versions of the network, we calculated the NMI of the membership and its respective shuffled version, and the obtained result is about $0.4$, which indicates that the shuffled values seem not to explain the obtained NMI values.

\subsection{Complementary strategies for defining modified networks}
\label{sec:strategies}

As a complement of the results in the previous section, we investigated the impact of more than one keyword deletion when creating the modified networks. Because there are many possible combinations of keywords, we considered two greedy-based strategies aimed at finding the best and worst scenarios in terms of the network sizes. These strategies are described as follows:
\begin{itemize}
   \item \emph{Best case}: for each iteration, a keyword is deleted from the original set of keywords, and the network is created. For choosing the keyword to be deleted, we employ a greedy-based strategy that searches for the smallest network. As adopted in the previous section, the network size is measured by considering the weakly connected component;
   \item \emph{Worst case}: the same as the Best case method, but for each iteration the greedy-based strategy searches for words that create the network with the highest number of nodes. 
\end{itemize}

\begin{figure*}[!ht]
  \centering
    \subfigure{\includegraphics[width=1.\textwidth]{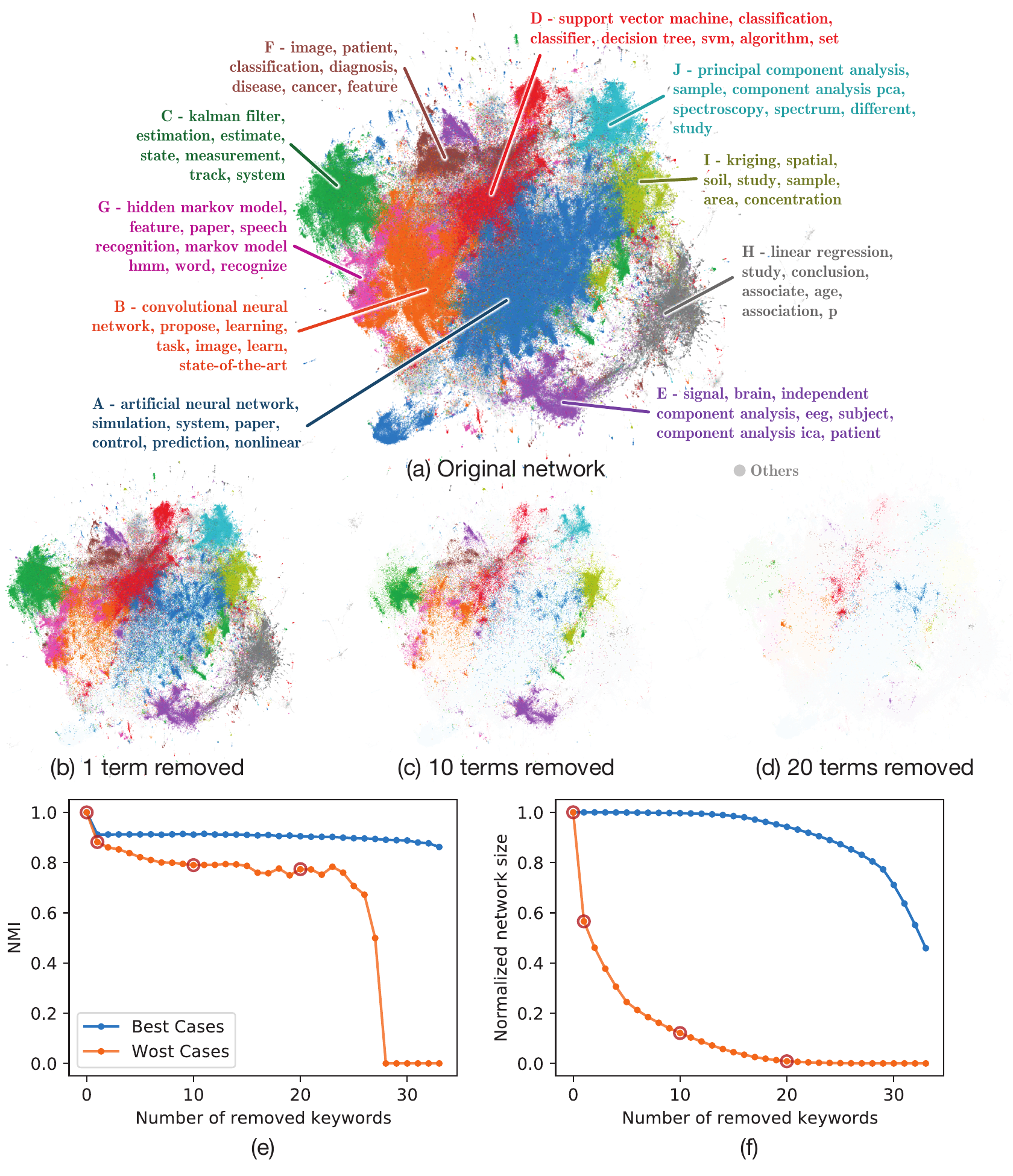}}
  \caption{(a) Original network visualization with the detected communities, and the respective subjects inferred by using the method proposed in~\cite{silva2016using}. Items (b), (c), and (d) show the modified network after removing 1, 10, and 20 terms, respectively. The measured NMI and the network sizes obtained for the best and worst case scenarios, are displayed in (e) and (f). The node positions were obtained by computing node2vec~\cite{grover2016node2vec} and UMAP~\cite{mcinnes2018umap} projection.}
  \label{fig:atack}
\end{figure*}

Figure~\ref{fig:atack}(a) shows the original network visualization, and Figures~\ref{fig:atack}(b)~(c)~and~(d) provide examples of modified networks by considering the worst case scenario. The NMI and the network sizes obtained from both strategies are shown in Figure~\ref{fig:atack}(e) and (f), respectively. In the best case, it is necessary to remove almost all keywords to reduce the network size to half of the original network. Moreover, for all cases, the measured NMI indicates that there is good agreement between the communities of the original and modified networks. The smallest network refers only to the keyword \emph{Neural network}. On the other hand, in the worst case, in the first step, removing a single keyword, \emph{Neural network}, strongly decreased the network size (see Figure~\ref{fig:atack}(b)). However, even with modified networks being much smaller than the original network, it is necessary to remove a substantial number of nodes to compromise the agreement between the original and the modified versions of the network.

The obtained results indicate that the employed network can be understood as being robust in terms of its communities. Furthermore, this finding suggests the two following hypotheses: (i) the communities did not substantially change when nodes were removed, and (ii) almost entire communities are left out when the keywords are removed. The number of detected communities is much more substantial than the number of employed keywords, with $22,163$ communities. Furthermore, on average, each community most frequent keyword represents $(76.15 \pm 21.46)\%$ of its nodes. So, a third hypothesis is that both (i) and (ii) happen, though with different intensities. However, we believe that hypothesis (i) better explains the results. This effect is further investigated in the next section by considering a toy model.

\subsection{Toy model}
\label{sec:res_toy}

\begin{figure}[!ht]
  \centering
    \subfigure{\includegraphics[width=0.45\textwidth]{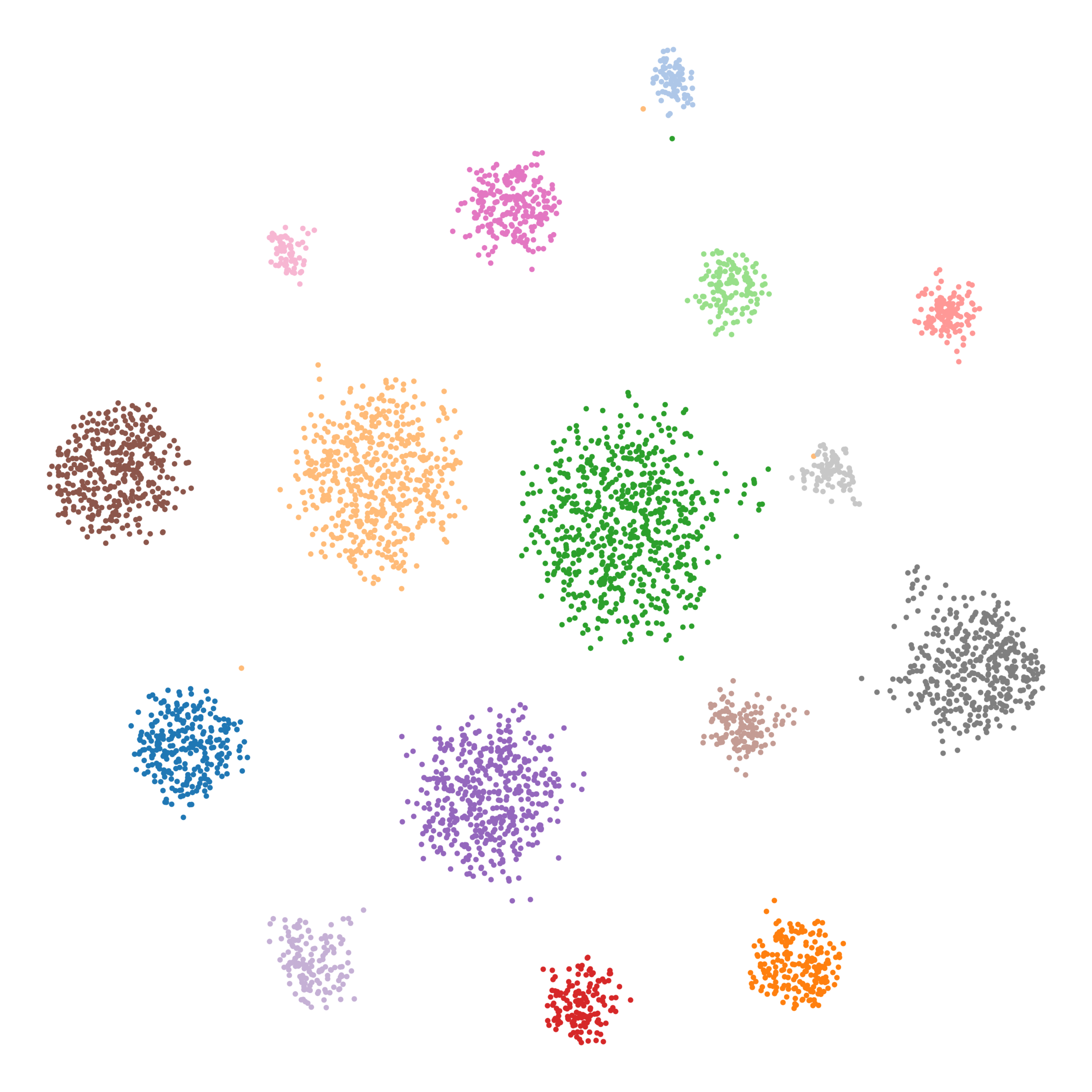}}
  \caption{Example of the visualization of a LFR-Benchmark network with $4000$ nodes. The node colors represent the communities generated by the LFR-Benchmark. The node positions were obtained by computing node2vec~\cite{grover2016node2vec} and UMAP~\cite{mcinnes2018umap} projection.}
  \label{fig:visualization_toy}
\end{figure}

In order to better understand the results observed in the previous section, we considered a toy model. For the sake of simplicity, we opted for having a single keyword defining the nodes since the vast majority of the nodes of the real network were obtained from a single keyword. Three different network sizes were considered:  2000, 4000, and 8000 nodes. The adopted parameters, described in Section~\ref{sec:toy}, are shown in Table~\ref{tab:par}. These parameters were empirically chosen in order to create networks with well-defined communities. An example of the employed networks is shown in Figure~\ref{fig:visualization_toy}.

\begin{table}[!ht]
\begin{tabular}{|c|c|c|c|c|c|c|c|}
\hline
\large Size & \large $\mu$ & \large $\gamma$ & \large $S_{min}$ & \large $S_{max}$ & \large  $k_{max}$ & \large $\langle k \rangle$ & \large $C$ \\ \hline \hline
\large 2000 & \large 0.15  & \large -2 & \large 50 & \large 1000 & \large 200 & \large 20 & \large 8 \\ \hline
\large 4000 & \large 0.15  & \large -2 & \large 50 & \large 1000 & \large 400 & \large 20 & \large 16 \\ \hline
\large 8000 & \large 0.15  & \large -2 & \large 50 & \large 1000 & \large 800 & \large 20 & \large 32 \\ \hline
\end{tabular}
\caption{Employed parameters for the LFR-Benchmark.}
\label{tab:par}
\end{table}

\begin{figure*}[!ht]
  \centering
    \subfigure[2000 nodes]{\includegraphics[width=0.32\textwidth]{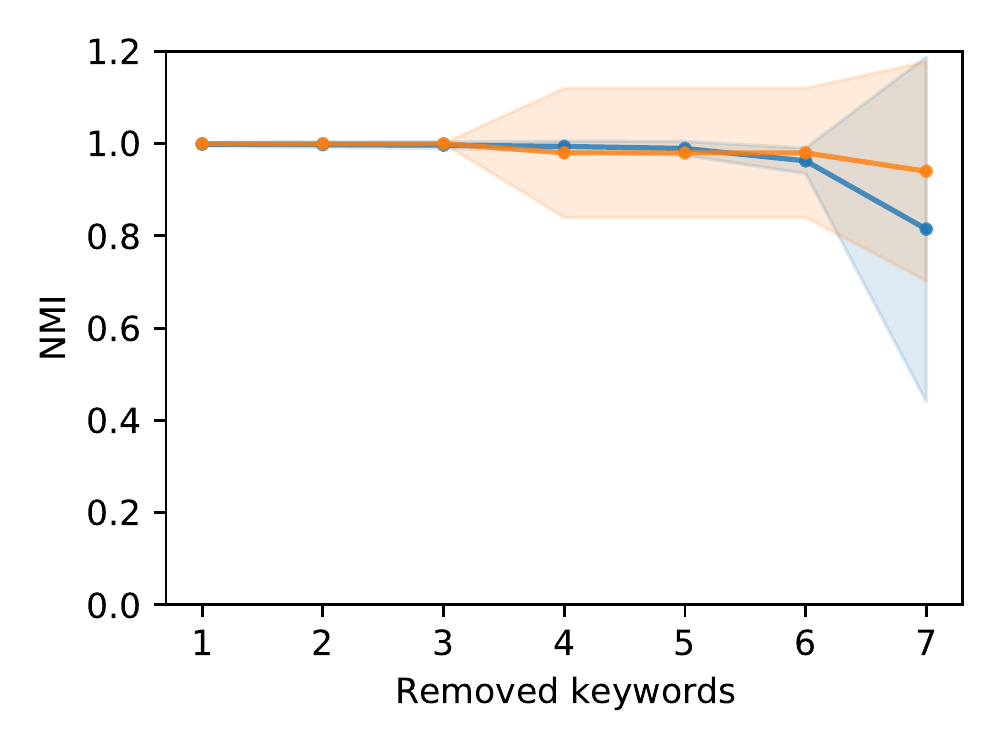}}
    \subfigure[4000 nodes]{\includegraphics[width=0.32\textwidth]{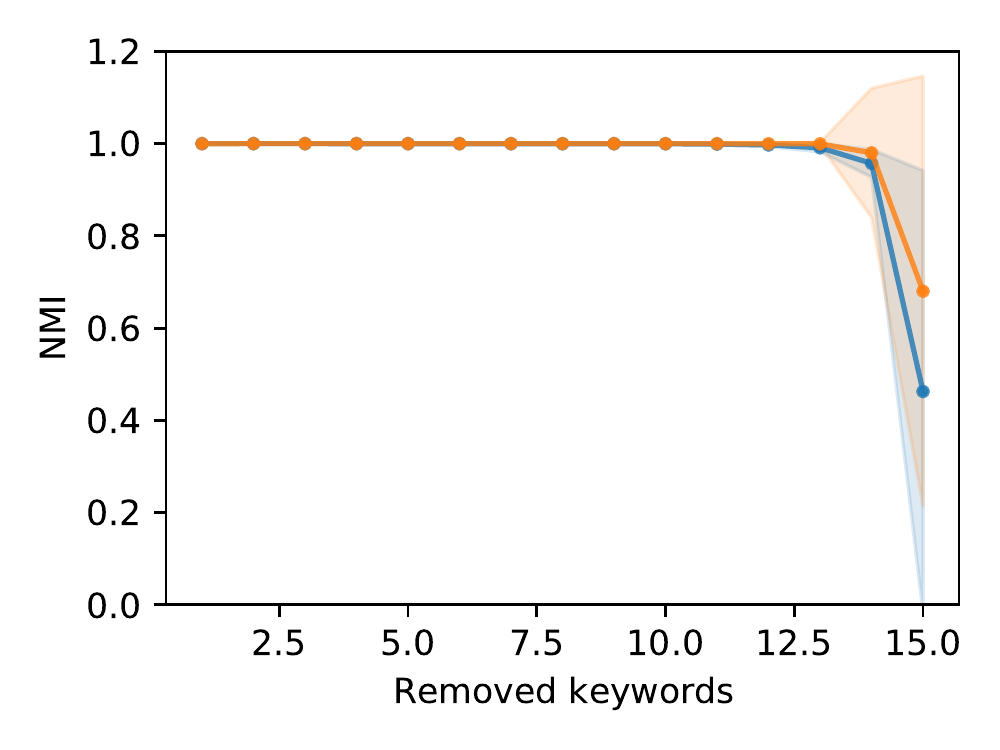}}
    \subfigure[8000 nodes]{\includegraphics[width=0.32\textwidth]{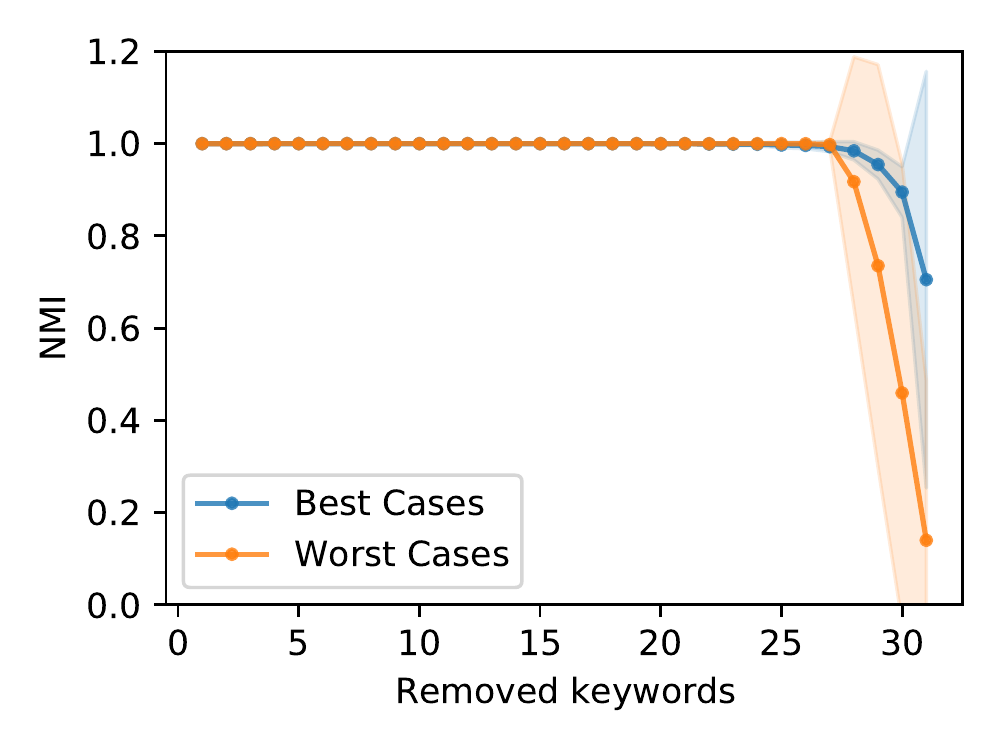}}
  \caption{Comparison between best and worst cases, by considering the LFR-Benchmark with different sizes. Here, the keywords are dependent of the communities. The shaded regions represent the standard deviations.}
  \label{fig:toymodel_attack_community}
\end{figure*}

As in Section~\ref{sec:strategies}, we consider both worst and best case strategies of node deletion. We start our analysis by considering the keywords as being dependent of the communities (described in Section~\ref{sec:toy}). Figure~\ref{fig:toymodel_attack_community} illustrates the obtained results for the three employed network sizes. The results presented in this section represent averages of $50$ executions. Interestingly, there was no significant difference between the best and worst-case scenarios. This result indicates that, even with the removal of entire communities, the remaining communities were correctly identified. Furthermore, the size of the removed community played a small role in the obtained results. The differences between the best and worst cases were observed only for the latest iterations.

\begin{figure*}[!ht]
  \centering
    \subfigure[2000 nodes]{\includegraphics[width=0.32\textwidth]{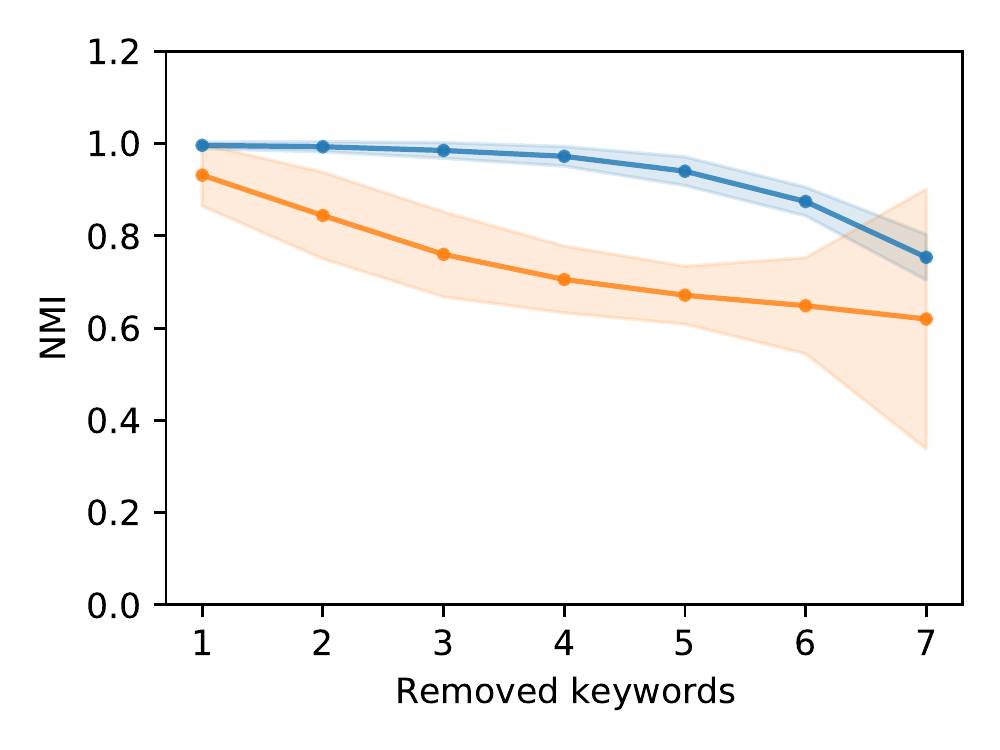}}
    \subfigure[4000 nodes]{\includegraphics[width=0.32\textwidth]{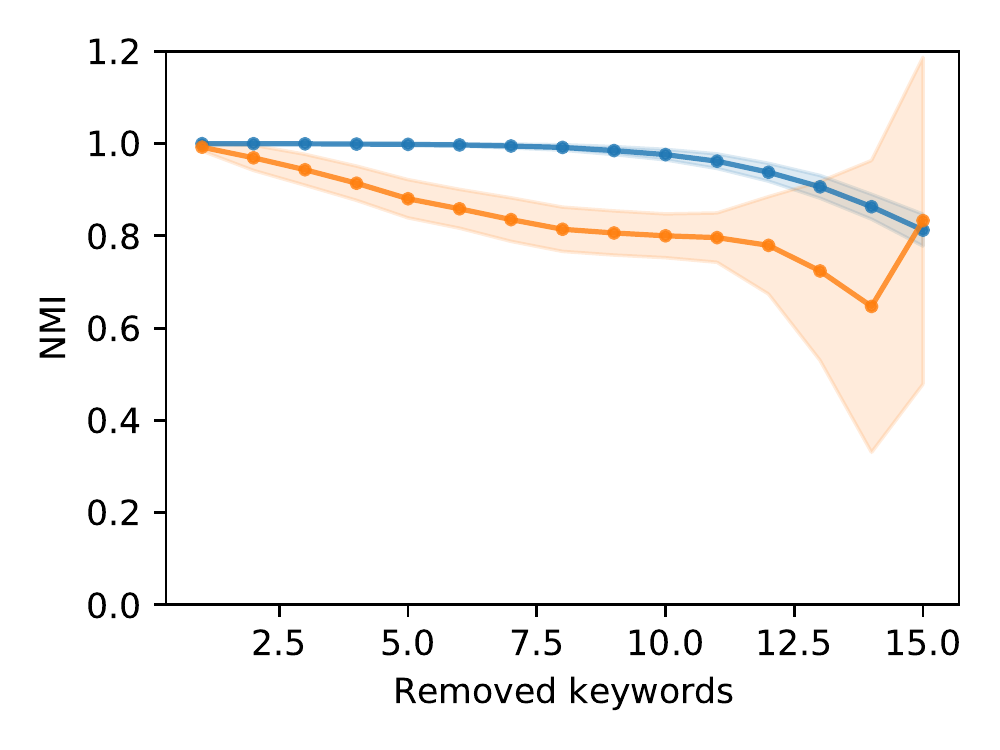}}
    \subfigure[8000 nodes]{\includegraphics[width=0.32\textwidth]{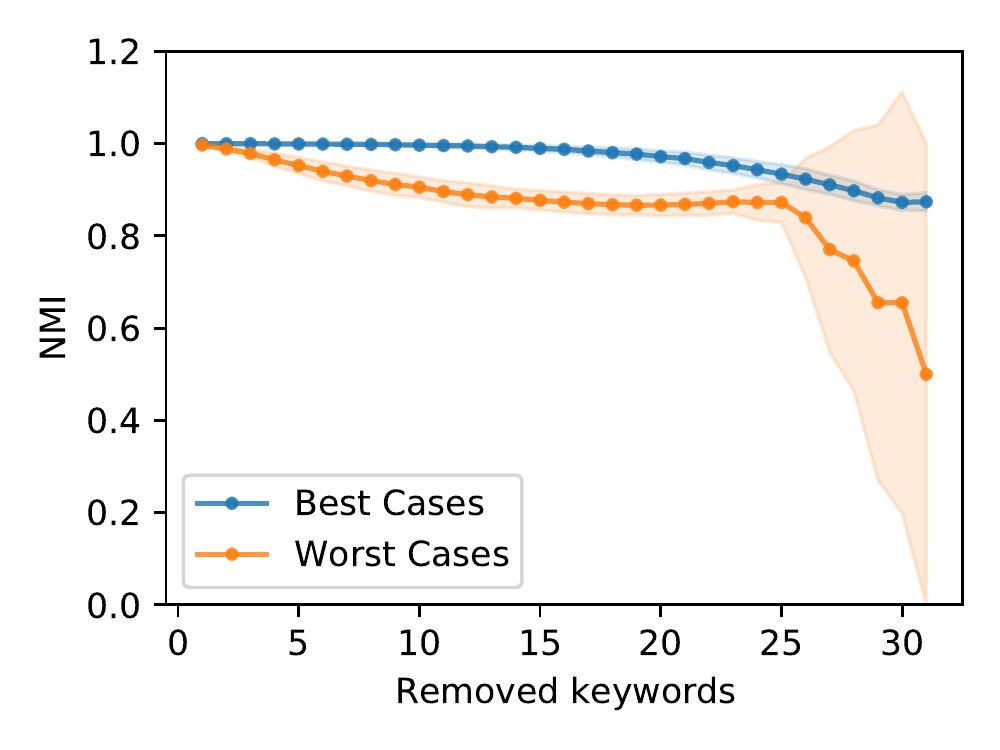}}
  \caption{Comparison between best and worst cases, by considering the LFR-Benchmark with different sizes. Here, the keywords are independent of the communities. The shaded regions represent the standard deviations. }
  \label{fig:toymodel_attack_random}
\end{figure*}

Next, we analyzed the strategy of keywords independent of communities (see Section~\ref{sec:toy}), as shown in Figure~\ref{fig:toymodel_attack_random}. First, there are noticeable differences between the results obtained from best and worst cases, which can be observed from the first removed keywords. In this case, the order of node deletions controls only the number of removed nodes. So, by removing more nodes, the network community organization was strongly influenced.

\begin{figure*}[!ht]
  \centering
    \subfigure[2000 nodes]{\includegraphics[width=0.32\textwidth]{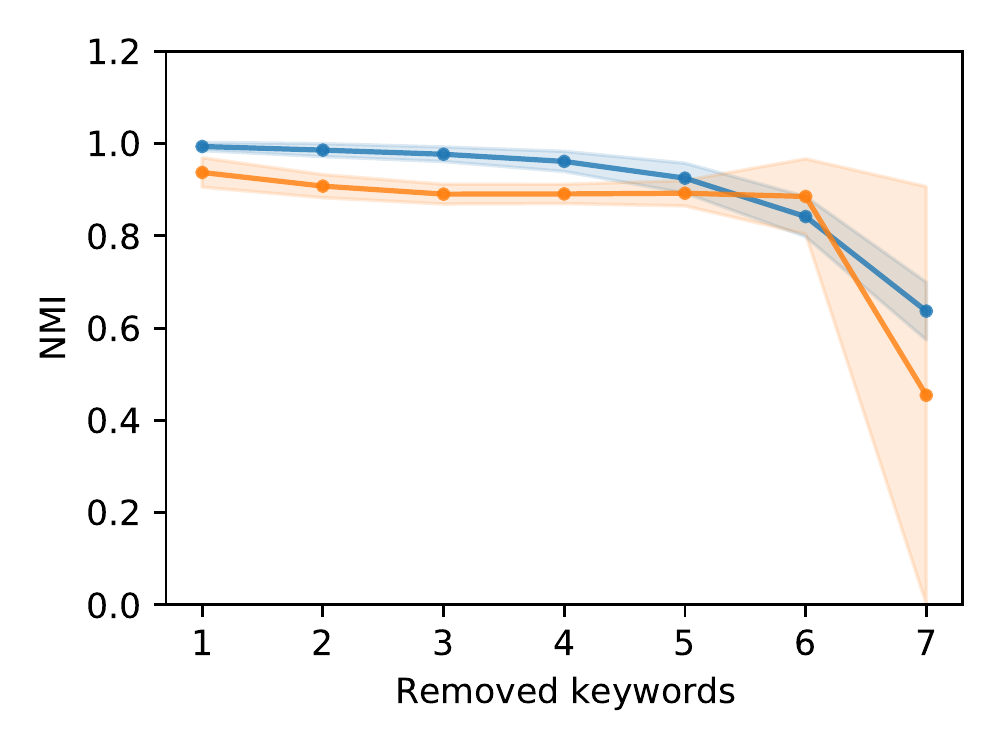}}
    \subfigure[4000 nodes]{\includegraphics[width=0.32\textwidth]{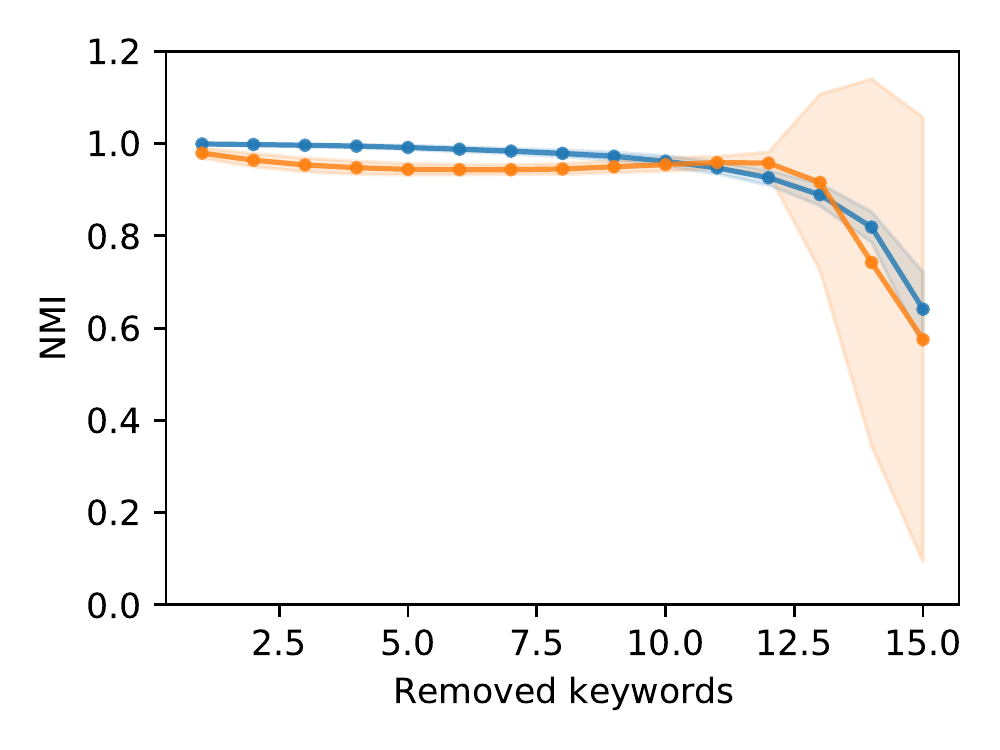}}
    \subfigure[8000 nodes]{\includegraphics[width=0.32\textwidth]{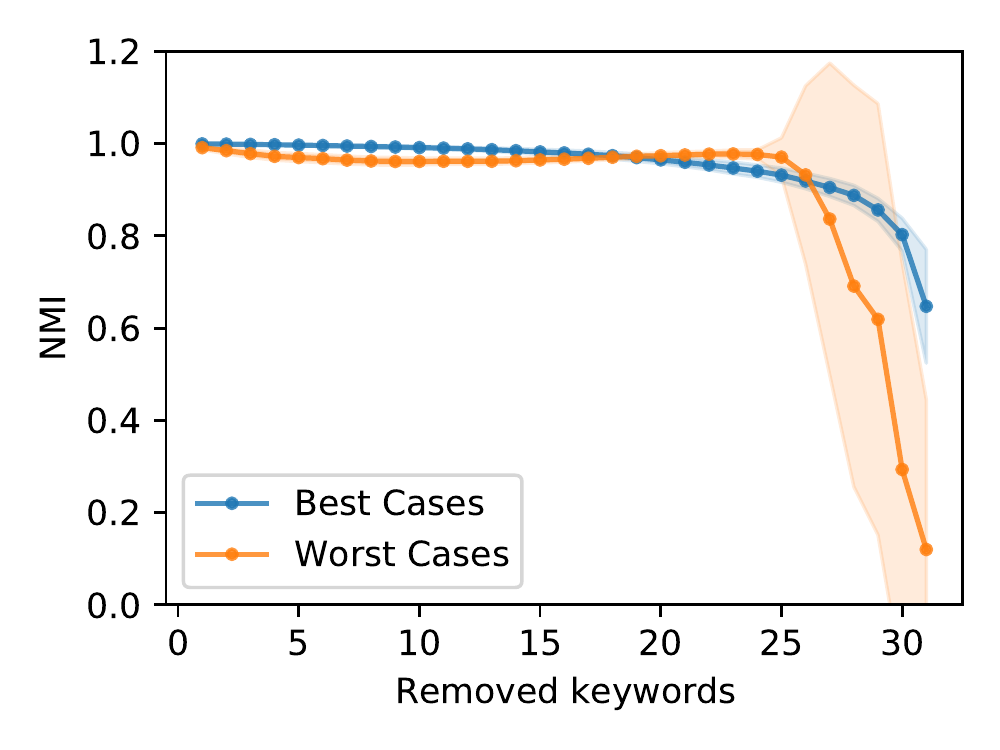}}
  \caption{Comparison between best and worst cases, by considering the LFR-Benchmark with different sizes. Here, a fraction of the keywords randomly shuffled between communities, which represents an intermediate scenario between the results of Figures~\ref{fig:toymodel_attack_community}~and~\ref{fig:toymodel_attack_random}. The shaded regions represent the standard deviations. }
  \label{fig:toymodel_attack_percentage}
\end{figure*}

In order to contrast the previously considered scenarios, we tested an intermediate possibility. In this case, we started with the keywords dependent on communities and shuffled a given percentage $\rho$ of these keywords. The employed parameter $\rho$ corresponds to that of the real network. More specifically, the most frequent keyword of a community represents, on average, $76.15\%$ of their nodes. So, we took $\rho$ as the complement of this value, i.e.,~ $\rho=0.2385$.
Figure~\ref{fig:toymodel_attack_percentage} illustrates the obtained results. As expected, the differences between the best and worst cases were found to be intermediate between the results shown in Figures~\ref{fig:toymodel_attack_community}~and~\ref{fig:toymodel_attack_random}. Thus, $\rho$ can simulate scenarios between the two proposed strategies.

The results described in this section illustrate that the toy model can reproduce some characteristics of the citation network. Furthermore, it paves the way to a more extensive range of scenarios. Interestingly, here we show that by varying the topology and employed parameters, the real scenario behavior is kept. For all of the considered parameter sets, the community organizations were found to be robust even when the majority of the nodes were removed. However, it was also found that the LFR-benchmark cannot perfectly represent the real network.  In this model, all communities tend to be connected. So, if we delete all nodes from a given community, the connected component is likely not to fragment.

\section{Conclusions} \label{sec:conc}

Interesting information can be obtained from citation networks, including the analysis of the advancement of science and the relationship between areas. One characteristic intrinsically related to these networks is the presence of communities, which can be interpreted as partitions into subareas. Consequently, from these communities, much information can be obtained, such as the identification of labels that characterize the content of each group. In~\cite{silva2016using}, the authors proposed a pipeline that starts with queries for data retrieval and yields the network communities and their respective identified labels. 

Since the community structure of citation networks has been frequently employed in many different studies, it is necessary to better understand how robust these networks are. For this analysis, we considered variations of the initial query with different sets of keywords. The obtained networks, modified versions, were compared with the networks acquired from the original keywords. We considered queries removing a single keyword, which were compared by employing some network measurements. We also estimated the network differences in terms of their community organization by using measurements for quantifying the quality of the obtained clusters. We also considered a toy model in order to complement our analysis. More specifically, by employing this model, it is possible to vary the parameters and test a more extensive range of network topologies.

While comparing among the network topologies, we found that the networks tend to be similar for most cases. However, for the network without considering the keyword ``Neural network'',  the size was substantially reduced, and other features were found to be markedly different from the other networks. By considering the original and the modified networks employed in the previous test, we compared the community structures. We considered many measurements of cluster quality. Interestingly, the network that varied the most was the network without the keyword ``Neural network'', which is the same in the previous test.  As a complementary analysis, we considered two strategies for defining the modified networks: best and worst cases. In the former, we removed keywords while keeping the network as large as possible. In the latter, the keywords were removed aiming at keeping the network as small as possible. For both analyses, we found that even with small networks, the community structure tends to be maintained.  Even though there are much more communities than employed keywords, we found that the latter tend to be intrinsic and stably related to the communities. More specifically, the majority of the documents of a given community could then be retrieved from the same keyword. In order to better understand how the citation networks undergo changes as the keywords are modified, a toy model was employed, which confirmed that the community structure tends to be kept for an ample range of network parameters. This experiment confirms that if the keywords are more dependent on specific communities, the community structure tends to be held independently of the employed strategy of modified versions.

The results found in this study pave the way to many related works. 
In this paper, we considered areas as recovered by a large set of keywords. A possible complementary study could contemplate the analysis of the robustness of communities on networks retrieved from  smaller sets of keywords. Another possibility is to investigate strategies to ensure or improve the robustness of these networks, such as by expanding the query terms. A toy model that considered more features based on the citation networks could also be considered.

\ 

\section*{Acknowledgments}
Alexandre Benatti thanks Coordenação de Aperfeiçoamento de Pessoal de Nível Superior - Brasil (CAPES) - Finance Code 001.
Henrique F. de Arruda acknowledges FAPESP for sponsorship (grant \#2018/10489-0).  César H. Comin thanks FAPESP (grant \#18/09125-4) for sponsorship.  Luciano da F. Costa thanks CNPq (grant \#307085/2018-0) and NAP-PRP-USP for sponsorship. This material is partially supported by AFOSR \#FA9550-19-1-0391.  This work has been supported also by FAPESP grants  \#2015/22308-2.

\bibliography{ref}

\end{document}